\documentclass[sigconf]{acmart}
\usepackage{booktabs} 
\usepackage{array}

\usepackage{cellspace}
\newcommand\cincludegraphics[2][]{\raisebox{-0.42\height}{\includegraphics[#1]{#2}}}

\usepackage{xspace}

\setcopyright{rightsretained}

\acmDOI{10.475/123_4}

\acmISBN{123-4567-24-567/08/06}

\acmConference[ICMR'17]{ACM International Conference on Multimedia Retreival}{June 6}{Bucharest, Romania} 
\acmYear{2017}
\copyrightyear{2017}

\acmPrice{15.00}

\begin{document}
\title[GANs for Multimodal Representation Learning in Video Hyperlinking]{
Generative Adversarial Networks for Multimodal Representation Learning in Video Hyperlinking}

\author{Vedran Vukoti{\'c}}
\affiliation{%
  \institution{INSA Rennes, INRIA/IRISA}
  \streetaddress{Rennes, France}
}
\email{vedran.vukotic@irisa.fr}

\author{Christian Raymond}
\affiliation{%
  \institution{INSA Rennes, INRIA/IRISA}
 \city{Rennes} 
  \state{France} 
}
\email{christian.raymond@irisa.fr}

\author{Guillaume Gravier}
\affiliation{%
  \institution{CNRS, INRIA/IRISA}
  \streetaddress{Rennes, France}
}
\email{guillaume.gravier@irisa.fr}



\begin{abstract}
Continuous multimodal representations suitable for multimodal information retrieval are usually obtained with methods that heavily rely on multimodal autoencoders. In video hyperlinking, a task that aims at retrieving video segments, the state of the art is a variation of two interlocked networks working in opposing directions. These systems provide good multimodal embeddings and are also capable of translating from one representation space to the other. Operating on representation spaces, these networks lack the ability to operate in the original spaces (text or image), which makes it difficult to visualize the crossmodal function, and do not generalize well to unseen data.
Recently, generative adversarial networks have gained popularity and have been used for generating realistic synthetic data and for obtaining high-level, single-modal latent representation spaces. In this work, we evaluate the feasibility of using GANs to obtain multimodal representations.
We show that GANs can be used for multimodal representation learning and that they provide multimodal representations that are superior to representations obtained with multimodal autoencoders. Additionally, we illustrate the ability of visualizing crossmodal translations that can provide human-interpretable insights on learned GAN-based video hyperlinking models.
\end{abstract}

%
%


\copyrightyear{2017}
\acmYear{2017}
\setcopyright{acmlicensed}
\acmConference{ICMR '17}{}{June 6--9, 2017, Bucharest, Romania} 
\acmPrice{15.00}
\acmDOI{DOI: http://dx.doi.org/10.1145/3078971.3079038}
\acmISBN{ACM ISBN 978-1-4503-4701-3/17/06}



\maketitle

\pagestyle{empty} 

\section{Introduction}

Automatic generation of hyperlinks in video collections presents a problem of growing interest, where video segments are interlinked by their high level semantic similarity, based on one or more of their modalities. Video hyperlinking consists of establishing links between relevant video segments within a collection. Links are defined between source video segments, called anchors (video segments that the users are viewing and query for additional relevant video segments), and target video segments (video segments relevant for to each queried anchor). For every anchor, a multimodal retrieval task is performed and a set containing the most relevant related targets is returned.
Automatic generation of hyperlinks within large video collections has been a subject studied and evaluated yearly as part of the MediaEval and TRECVID initiatives, within their Search and Hyperlinking benchmark~\cite{awad2016trecvid}.
There are different approaches to handling multiple modalities in
multimodal retrieval and thus in video hyperlinking.  Straightforward
approaches using multiple modalities consist either in comparing each
modality separately, combining the scores or rankings, or in combining
descriptors from each modality. These methods however do not fully
exploit the multimodality of the datasets. Regardless of classical
methods such as canonical correlation analysis (CCA), the state of the
art methods yielding a joint multimodal representation space from the
initially disjoint representations heavily rely on multimodal
autoencoders
~\cite{feng2014cross,cha2015multimodal}.

Multimodal AEs operate with representations of each modality (in our case, speech and visual representations). It is possible to translate from one initial representation space to the other, but not completely to the original space. Being able to visualize the crossmodal mappings (e.g., what images are expected given a certain speech transcript) in video hyperlinking can offer important human-interpretable insight about the learned model and explain its performance. Recently, generative adversarial networks~\cite{goodfellow2014generative} have become popular as means of generating realistic artificial data in the original, initial space (as opposed to the original representation space). Even more recently, generative adversarial autoencoder models have been conditioned on the input with one modality (CGAN), while allowing them to generate realistic artificial data of the other modality~\cite{reed2016generative} thus allowing for crossmodal translation from one initial space (e.g., text) to the other initial space (e.g., image), as opposed to crossmodal translation between representation spaces. We explore the possibility of using CGANs in video hyperlinking and creating models that not only provide good multimodal embeddings but also provide human-interpretable visualizations of what the model has learned.

Multimodal AEs are trained in an unsupervised manner on a dataset of finite, often limited size. It is possible to enlarge the dataset by applying random transformations and noise to the existing samples but the amount of samples will sill remain limited. In this work, we explore the possibility of using CGANs to tackle this problem. A generator network can generate an unlimited number of synthetic samples to train with and can constantly be improved by competing with the discriminator network. The discriminator network can also benefit from the additional synthetic samples and provide even better multimodal embeddings.

\begin{table*}[!htb]
\centering
\caption{Visualization of generated synthetic images given automatic transcripts (in lowercase, with some stop words removed) as input. In the last column, a real image from the video segment corresponding to the input automatic transcript is shown.
CGANs provide good visualizations of the video hyperlinking model: in the last row, given speech transcripts related to war thematic, the model is expecting a news presenter, while the actual video segment contains footage from the war zone.}
\label{table:speech_to_images}
\begin{tabular}{|m{8cm}|Sc|Sc|}
\hline
\multicolumn{1}{|c|}{\textbf{Input - Automatic Speech Transcript}} & \textbf{Generated Images} & \textbf{Real Image} \\ \hline \hline

``\ldots 
insects emerged to take advantage of the abundance . the warm weather sees the arrival of migrant birds stone chests have spent the winter in the south  \ldots''
& \cincludegraphics[scale=0.5]{"images/nature"} 
& \cincludegraphics[scale=0.5]{"images/nature_real"}
\\ \hline

``\ldots 
second navigation of the united kingdom . the north sea , it was at the north yorkshire moors between the 2 , starting point for the next leg of our journey along the coast \ldots''
& \cincludegraphics[scale=0.5]{"images/water"} 
& \cincludegraphics[scale=0.5]{"images/water_real"}
\\ \hline    


``\ldots  the role of my squadron afghanistan is to provide the the reconnaissance capability to use its or so forgave so using light armor of maneuvering around the area of 
\ldots''
& \cincludegraphics[scale=0.5]{"images/news"} 
& \cincludegraphics[scale=0.5]{"images/news_real"}
\\ \hline

\end{tabular}
\end{table*}

\section{Related Work}
Multimodal AEs are commonly deployed in tasks where multiple modalities are used.
Such tasks include but are not limited to classification~\cite{cha2015multimodal, ngiam2011multimodal} and retrieval
~\cite{feng2014cross, lu2015semantic}. More specifically, in video hyperlinking, the current state of the art is a variation on multimodal AEs, consisting of two interlocked crossmodal DNNs, translating in opposing directions~\cite{vukotic-icmr-16, vukotic-ivlmm-16} that achieved the best performance at the latest TRECVID evaluation~\cite{irisa2016trecvid}.

Generative adversarial networks have been first introduced by~\cite{goodfellow2014generative} as a two-network generative-discriminative model 
for generating high-quality, single-modal, realistic samples that could be mistaken for real samples from the dataset the model is trained on. GANs quickly became popular and are now used in a multitude of tasks, such as generating super-resolution images, 
inpainting, 
de-occlusion 
and many others. Conditional GANs~\cite{mirza2014conditional} have been shown to generate realistic samples of one modality, given a conditioning input of another modality. A typical example of a crossmodal CGAN model is text to image synthesis~\cite{reed2016generative, zhang2016stackgan} but the conditioned input is not necessarily bound to multiple modalities and can be conditioned also on the same modality.

For the purpose of multimodal retrieval, we focus on crossmodal / multimodal CGANs and ways to obtain meaningful multimodal representations from them. While multimodal setups are currently less explored, there is a lot of evidence that GANs learn meaningful representations in single-modal setups
~\cite{makhzani2015adversarial, chen2016infogan, perarnau2016invertible},
most notably in the generator network. These representations are obtained in a completely unsupervised manner and can be used to model changes in style, pose, 
color
or even style and structure in RGBD data.
Evidence suggests that both the generator network and the discriminator network can produce meaningful single-modal representations~\cite{radford2015unsupervised}. Our work falls into the category of text to image CGANs~\cite{reed2016generative, zhang2016stackgan}, where we explore the possibility of obtaining good multimodal representations from the discriminator, while using the generator to visualize crossmodal mappings for the purposes of multimodal retrieval in video hyperlinking and to improve the embeddings from the discriminator network.


\begin{figure}[tb]
    \centering
	\includegraphics[scale=0.3]{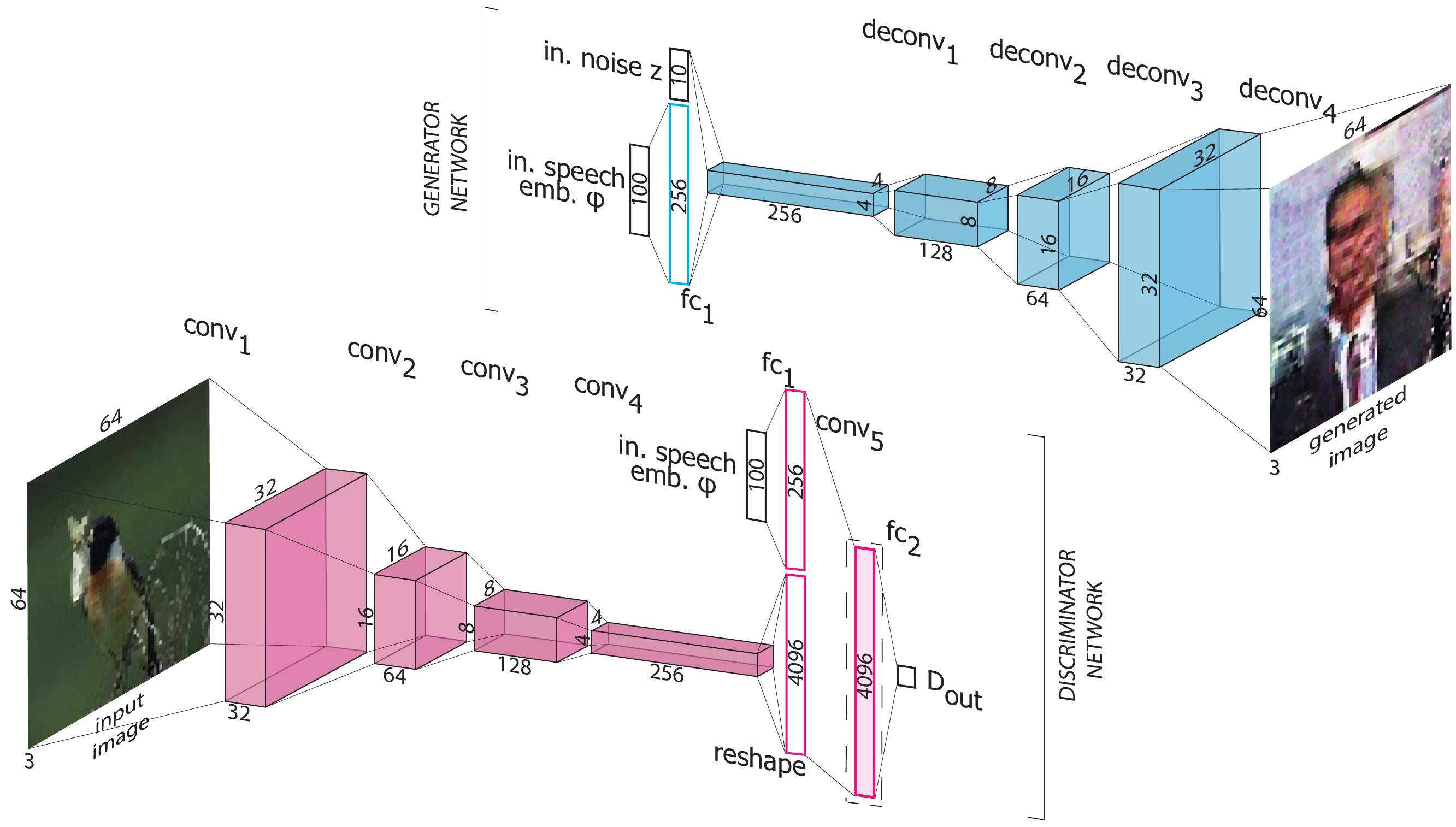}
	\caption{Used CGAN architecture, consisting of a generator and a discriminator network. Multimodal embeddings are obtained in the discriminator, after the last, 1D convolution operation, as denoted with a dashed rectangle.}
    \label{fig:arch}
    \vspace{-7mm}
\end{figure}

\section{Methodology}
We split the work in two separate parts. In the first part, we evaluate the current state of the art multimodal AEs. In the second part, we evaluate the feasibility of using CGANs to perform multimodal embedding and we compare the resulting representations against the current state of the art in the setup of multimodal retrieval within the video hyperlinking domain.

\subsection{Multimodal Autoencoders}
Multimodal AEs with two separate input and output bra\-nches~\cite{ngiam2011multimodal}, one for each modality have been shown to perform better than simple AEs where the two modalities are concatenated. Recently, BiDNN, a variation with two disjoint but tied networks, working in opposing directions has been shown to define the new state of the art in video hyperlinking~\cite{irisa2016trecvid}. We compare our proposed model against both classical multimodal AEs and the recent BiDNN variation. However, multimodal AEs need existing continuous representations as inputs. We opt for averaged Word2Vec for representing speech and VGG-19 features for representing video keyframes.

\subsection{Unsupervised Multimodal Representation Learning with CGANs}
\label{gan_theory}
The CGAN architecture we evaluate is based on the GAN-CLS text to image model~\cite{reed2016generative}. The model consists of a generator network $G$ and a discriminator network $D$. The generator network takes a noise vector $z$, sampled from a $\mathcal{U}(-1, 1)$ distribution and a text embedding $\varphi$, and generates a synthetic image $\hat{x} = G(z, \varphi)$. The generator consists of a separate fully-connected layer with a leaky-ReLU activation function (that allows for a small non-zero gradient on the negative side) for modeling the text embeddings. This is then concatenated with the noise input. A standard deconvolutional network follows. The discriminator network takes an image $x$ and a text embedding $\varphi$ and determines whether the pair is real or artificial $D(x, \varphi)$. Contrary to the basic CGAN model that only differentiates between real and synthetic images, CGAN-CLS is trained on three pairs. In our case \{real image, real text\}, \{incorrect image, real text\} and \{synthetic image, real text\}. As we are doing unsupervised learning, pairs with a non-matching image and text are chosen at random from different video segments and the dataset is not split. Pairs of real matching and non-matching text-image pairs are necessary to train the discriminator. The discriminator consists of a series of convolutional layers with batch normalization, followed by a leaky-ReLU for modelling the input image and a full-connected leaky-ReLU layer to model the text embeddings. The two branches are then concatenated and a $1 \times 1$ convolution is performed, followed by batch normalization and a leaky-ReLU activation before obtaining the discriminator score. For multimodal embedding, we use the vector obtained after the final $1\times 1$ convolution of the discriminator. To maintain the losses of both the generator and the discriminator at similar levels (having a discriminator that is performing too well would prevent the generator from converging), the generator is updated four times more often than the discriminator. A cosine distance between the obtained multimodal embeddings is used to measure the similarity of the desired video segments both in the case of multimodal AEs and CGANs.


\section{Experiments}
In this section we first define the dataset and the specific hyperparameters of each architecture and then analyze and discuss the results of each experiment.

\subsection{Dataset and Single-Modal Representations}
\label{dataset}
We test all the methods in a video hyperlinking setup on a BBC video hyperlinkig dataset with segments and the groundtruth obtained post-evaluation during MediaEval 2014. The task considered is that of ranking target segments given a source segment acting as a query. We use only the video segments that contain both speech transcripts and video keyframes (10,321 in total). Speech is represented with averaged embeddings of size 100 from a skip-gram \textit{Word2Vec} model with hierarchical sampling and a window size of 5~\cite{irisa2016trecvid}. Regarding the visual stream, a representative keyframe is chosen for each video segment by computing a histogram for each pixel over all the keyframes and choosing the keyframe closest to the median
as a representative of each short video segment. Such an image is then downsampled and cropped to a size of $64 \times 64$ pixels. To obtain a sensible comparison, the same data and image sizes are used for all the methods. The CGAN model works directly with the text embeddings and images. Multimodal AEs also need visual embeddings as inputs. We obtained them from the same images, with a VGG-19 network (pretrained on ImageNet) and they are of size 4096. Table~\ref{table:embeddings} show the initial single-modal results, which are 56.55\% for speech transcripts and 52.41\% for video keyframes, in terms of precision at 10~\cite{awad2016trecvid}.

\begin{table}[!htb]
\centering
\caption{Comparison of initial modal single-modal representations and multimodal embeddings obtained with different methods. For each method, precision at 10 and its respective standard deviation are reported.}
\label{table:embeddings}
\begin{tabular}{|l|c|c|}
\hline
\textbf{Representation} & \multicolumn{1}{l|}{\textbf{$P_{10}$ (\%)}} & \multicolumn{1}{l|}{$\sigma$ (\%)} \\ \hline \hline
Speech Transcripts Only &              56.55                 & -                                    \\ \hline
Visual Only (VGG-19)    &              52.41                 & -                                    \\ \hline \hline
Multimodal AE           &              57.94                 & 0.82                               \\ \hline
BiDNN                   &              59.66                 & 0.84                                \\ \hline
\textbf{CGAN}           &              \textbf{62.84 }       & \textbf{1.36 }                      \\ \hline
\end{tabular}
\vspace{-5mm}
\end{table}

\subsection{Multimodal Autoencoders}
We evaluate both classical multimodal AEs and BiDNNs, their state-of-the-art variation in video hyperlinking. We chose an AE architecture with separate input and output branches of size 1000 for each modality  (first and last layers of consisting of two separate hidden layers of size 1000 each) and a common hidden layer or size 1000. For BiDNNs, we use the implementation provided by~\cite{irisa2016trecvid}. We use only the video segments that contain multimodal information (both speech transcripts and visual information) and we use averaged \textit{Word2Vec} embeddings for speech and VGG-19 embeddings for the keyframes, as defined in section~\ref{dataset}. BiDNNs expectedly outperform classical multimodal autoencoders and obtain a precision at 10 of 59.66\%.

\subsection{Multimodal Embedding with CGANs}
To evaluate the feasibility of obtaining multimodal embeddings with CGANs, we used the architecture described in Sec.~\ref{gan_theory} with a speech embedding of dimension 100, a noise input of size 10, a fully connected layer for modeling text of size 256 and 4 \textit{``deconvolutional''} layers of increasing size, each with 256, 128, 64 and 32 feature maps respectively in the generator network. The output of the generator is an image of size $64 \times 64$. The discriminator network consists of an image input of size $64 \times 64$ followed by 4 convolutional layers with batch normalization, each with 32, 64, 128 and 256 feature maps respectively and decreasing sizes and a speech embedding input of size 100, followed by a fully connected layer of size 256. The two branches are then concatenated and followed by another, final convolution.  We trained the network for 1000 epochs by using the Adam optimizer with a learning rate $r$ of $0.0001$, a momentum $\beta$ of $0.5$ and a batch size of $64$. As specified in Section~\ref{gan_theory}, we use the one-dimensional convolutional layer in the discriminator that follows the merging of the two branches, before batch normalization and activation as a multimodal embedding layer, as illustrated in Figure~\ref{fig:arch}. Other layers did not perform well and provided a lower or equal quality than the initial single-modal inputs.
The generator network does not only provide a mean to visualize crossmodal mappings but is also crucial for obtaining high-quality embeddings in the discriminator by competing with it and providing additional synthetic samples.

Compared to multimodal AEs, CGANs are computationally expensive to train (20h on a GPU, compared to a few hours on a CPU for BiDNN), even for small images, and require both modalities to be present. However, the discriminator of a CGAN provides multimodal embeddings that are greatly improved over the initial representation spaces or each modality. The results are shown in Table~\ref{table:embeddings}. Representations learned with a CGAN not only outperformed multimodal AEs but they also significantly ($p=99.9\,\%$ with a single-sided t-test) outperformed the state of the art BiDNN model and obtained 62.84\%.

\subsection{Crossmodal Visualizations}
The existence of a generator network is crucial to train a discriminator that provides well-performing multimodal embeddings. Interestingly, the generator network can also be used to visualize crossmodal mappings in video hyperlinking. The generator can straightforwardly create synthetic images given an embedding of the speech transcripts as input. Examples of images generated for real transcripts on a few video segments, as well as their real visual counterparts are displayed in Table~\ref{table:speech_to_images}.  The generator can also be reversed by simply applying the transposed learned kernels in an inversed  architecture and slicing the obtained vector. In this case, given an image, embeddings in the textual domain can be obtained. A few examples of that translation are shown in Table~\ref{table:images_to_text}.

\section{Conclusions and Future Work}
In this work we experimentally demonstrated the interest of CGANs as an alternative method to design multimodal embeddings. While CGANs are more computationally intensive to train than multimodal autoencoders, and are thus limited to smaller image sizes, they can provide better multimodal embeddings in the last layer of their discriminator network that outperform embeddings of multimodal autoencoders.
The generator network, that provides unlimited synthetic samples and is crucial for training the discriminator, can additionally be used to visualize one modality from another and provide human-interpretable insights of the crossmodal translations learned with this type of video hyperlinking models. To use CGANs for multimodal embedding more efficiently, it is necessary to investigate different architectures that are able to converge with bigger image sizes within a feasible training timeframe and evaluate possible methods of decreasing the training time for uses where online learning is required.

\section{Acknowledgments}
We gratefully acknowledge the support of NVIDIA Corporation with the donation of the Titan X GPU used for this research.

\begin{table}[!htb]
\centering
\caption{Visualization of the top words in the representation space of automatic transcripts, given an input image.}
\label{table:images_to_text}
\vspace{-3mm}
\begin{tabular}{|c|m{6cm}|}
\hline
\textbf{Input Image} & \textbf{Top Words in the Speech Modality} \\ \hline \hline
\cincludegraphics[scale=0.5]{"images/image_to_text_1"} &
britain, protecting, shipyard, carriers, jobs, vessels, current, royal, aircraft, securing, critics, flagships, foreclosures, economic, national
\\ \hline

\cincludegraphics[scale=0.5]{"images/image_to_text_2"} &
north, central, rain, northern, eastern, across, scotland, southwest, west, north-east, northeast, south, affecting, england, midlands
\\ \hline

\cincludegraphics[scale=0.5]{"images/image_to_text_3"} &
pepper, garlic, sauce, cumin, chopped, ginger, tomatoes, peppers,  onion,  crispy, parsley, grated, coconuts, salt, crust
\\ \hline


\end{tabular}
\vspace{-5mm}
\end{table}





\bibliographystyle{ACM-Reference-Format}
\bibliography{sigproc} 

\end{document}